\documentclass[pra,twocolumn,showpacs,showkeys]{revtex4}

\usepackage[latin1]{inputenc} 
\usepackage{graphicx}
\usepackage{amsmath}
\usepackage{color}

\begin{document}

\title{Adiabatic Passage and Spin Locking in Tm$^{3+}$:YAG}
\author{M. F. Pascual-Winter, R. C. Tongning, R. Lauro, A. Louchet-Chauvet, T. Chaneli\`ere, J.- L. Le Gou\"et}
\email[electronic mail: ]{jean-louis.legouet@lac.u-psud.fr}
\affiliation{Laboratoire Aim\'e Cotton, CNRS UPR3321, Univ. Paris Sud, b\^atiment 505,
campus universitaire, 91405 Orsay, France}

\begin{abstract}
In low concentration Tm$^{3+}$:YAG, we observe efficient adiabatic rapid passage (ARP) of thulium nuclear spin over flipping times much longer than $T_2$. Efficient ARP with long flipping time has been observed in monoatomic solids for decades and has been analyzed in terms of spin temperature and of the thermodynamic equilibrium of a coupled spin ensemble. In low-concentration impurity-doped crystals the spin temperature concept may be questioned. A single spin model should be preferred since the impurity ions are weakly coupled together but interact with the numerous off-resonant matrix ions that originate the spin-spin relaxation. The experiment takes place in the context of quantum information investigation, involving impurity-doped crystals, spin hyperpolarization by optical pumping, and optical detection of the spin evolution.
       
\end{abstract}




\maketitle
\section{Introduction}\label{intro}
For a decade or so, quantum information research has renewed interest in basic NMR processes, often correlated with optical excitation. For instance, quantum storage of light in impurity doped-crystals generally involves the conversion of optical coherence into spin coherence and back~\cite{fleischhauer2000,moiseev2001,nilsson2005,lauro2009a,afzelius2009,afzelius2010}. Indeed, information is preferably stored in the spin coherence that usually relaxes much more slowly than the optical dipole. The phase shift resulting from the inhomogeneous broadening of the spin transition is eliminated with refocusing techniques such as spin echoes~\cite{turukhin2001,longdell2005}. 

At low temperature, coherence relaxation is dominated by interaction with the fluctuating magnetic field generated by the crystal nuclear spins. Sensitivity to this field can be reduced by application of a properly sized and oriented external magnetic field. Indeed, by adjusting the energy splitting of the hyperfine transition to an extremum in three dimensions, one can make the first order Zeeman shift of the transition vanish~\cite{fraval2004,lovric2011}.  

Further control of transverse relaxation (TR) is obtained by dynamical decoupling with multiple rf pulse sequences~\cite{carr1954,meiboom1958,fraval2005,longdell2005}. Interaction with the environment gives rise to phase-shifts that are reversed periodically by the rf $\pi$-pulses. As a result, the phase-shifts cancel out as if the system were decoupled from external perturbations. 

Strong resonant excitation by a continuous rf field is also able to partly decouple the spins from their environment, with the restriction that only the magnetization component along the field enjoys the decoupling effect. Many NMR techniques have been dwelling on this spin-locking feature~\cite{solomon1959}, starting with the well known double nuclear resonance~\cite{hartmann1962}. Similar locking of electronic dipoles has been observed at optical frequency, both in solids~\cite{devoe1983,pryde2004} and gases~\cite{yodh1984,carlson1984,tchenio1988}. 

The spin-locking picture can be extended to adiabatic rapid passage (ARP) where the rf field is frequency swept through resonance with the spin transition and makes the magnetization flip together with the driving vector. 

The present paper reports on ARP on the I$=1/2$ nuclear spin of low concentration Tm$^{3+}$:YAG, an actively investigated system in the prospect of quantum storage~\cite{lauro2009a, deSeze2006, louchet2007, louchet2008, lauro2009, chaneliere2010,bonarota2010,lauro2011,bonarota2011}. We have been able to observe efficient ARP for various flipping time values, ranging from much less to much more than the transverse relaxation time T$_2$. This result is apparently consistent with the theory developed by Redfield~\cite{redfield1955,abragam1961,slichter1961,solomon1962} to overcome the failure of Bloch equations~\cite{bloch1946} on time scales larger than T$_2$. However this apparent agreement with textbook predictions might be misleading. Indeed Redfield description relies on spin temperature and spin thermodynamical equilibrium. Those notions apply to spin ensembles in monoatomic samples. The low concentration impurities we are considering here are too weakly coupled together to reach mutual equilibrium on the required timescale. In addition, we optically select a small fraction of the impurity ions, and optically pump their spin into $\pm$zero-temperature states, far from equilibrium with the environment. Hence we have preferred an alternative description, based on the radiation locking investigations that were conducted on optical transitions~\cite{schenzle1984,kessel1988}.    

In section~\ref{theory} we outline the difference between the single spin and the coupled spin ensemble approaches in the ARP context. Then we give a simple picture of dephasing inhibition in ARP. The experimental framework and results are examined in Section~\ref{experiment}.

\section{Adiabatic passage and dephasing inhibition}\label{theory}
\subsection{Fixed-frequency spin locking and frequency-swept adiabatic passage}
We first consider excitation by a fixed frequency rf field $\textbf{H}_1\exp(i\omega t)$, set orthogonal to the static field $\textbf{H}_0$. In the frame rotating around $\textbf{H}_0$ at angular speed $\omega$, the rf field is constant, $\textbf{H}_1$ being directed along axis $Ox$. Let the rf field be tuned to resonance with the spin transition at frequency $\omega_0$. In a typical spin-locking experiment~\cite{solomon1959}, the magnetization, initially aligned along $Oz$, is first rotated into direction $Oy$ by a $\pi$/2 rf pulse, much shorter than the inverse inhomogeneous distribution of the spin transition frequency. Then the rf field phase is abruptly changed by $\pi$/2, which rotates $\textbf{H}_1$ into the same $Oy$ direction as the magnetization. The rf field then holds the spins aligned along $Oy$. When strong enough, the locking field decouples the spins along $Oy$ from the environment. Therefore, the effective spin lifetime $T_{2y}$ raises from $T_2$ to $T_1$, that respectively denote the spin-spin and the spin-lattice relaxation times. However, only the spin component along $Oy$ enjoys such a lifetime increase. 

In an ARP experiment, one varies the rf field frequency continuously at rate $r$. Oscillating at instantaneous frequency $\omega(t)=\omega_0+rt$, the rf field goes through resonance with the spin transition at time $t=0$. In the frame rotating at angular speed $\omega(t)$, $\textbf{H}_1$ is directed along $Ox$. The magnetization is driven by vector $\mathbf{\Omega}=(\gamma H_1,0,rt)$, where $\gamma$ represents the gyromagnetic ratio. This vector flips upward or downward, depending on the sign of $r$, as time runs from $-\infty$ to $\infty$. Being initially aligned along $Oz$, the magnetization keeps locked to $\mathbf{\Omega}$, and flips along with it, provided $\mathbf{\Omega}$ rotates in the $xOz$ plane at rate slower than $\Omega$. In accordance with Bloch equations~\cite{bloch1946}, the magnetization should undergo transverse relaxation during the flipping time, when the detuning $|\omega(t)-\omega_0|$ gets smaller than $\gamma H_1$. This leads to the decay factor $\mathrm{exp}(-\tau_f/T_2)$~\cite{lacour2007}, where $\tau_f=\pi\gamma H_1/r$ represents the flipping time. Hence the magnetization size should be preserved provided $\tau_f\ll T_2$. However, quite in the same way as in fixed-frequency spin-locking, the magnetization is expected to be decoupled from the environment by locking to $\mathbf{\Omega}$, and the relevant relaxation time is expected to grow well beyond $T_2$. 
\subsection{Spin ensemble and single spin pictures}
At first sight we might feel satisfied with the above description. Decoupling from the environment is apparently consistent with Redfield prediction of spin-locking and adiabatic passage on time scales much longer than $T_2$. A closer look reveals critical discrepancies. Our reasoning relies on the separation of rf excitation and relaxation. We have implicitely assumed that each single spin interacts with the driving field and with a large reservoir, the latter being insensitive to the field. Usually such a picture does not hold in NMR. Instead the driving field simultaneously excites an ensemble of coupled spins, and spin-spin relaxation results from the coupling of those excited spins. The thermodynamic approach~\cite{redfield1955,abragam1961} has proven to deal correctly with this complex problem. On time scales shorter than $T_1$, the spin ensemble is a closed system, decoupled from the lattice, that reaches equilibrium within $T_2$. A spin temperature may be defined at each moment, provided the transformations proceed on a time scale longer than $T_2$. In ARP the spin system undergoes a reversible, isoentropic transformation as $\omega(t)-\omega_0$ varies, which maintains the magnetization aligned along the driving vector. In this picture, the magnetization is preserved when $T_2\ll\tau_f$, just at the opposite of the condition imposed by Bloch equations. This feature was used in the past to detect weak transverse magnetization, in materials such as silicon~\cite{abragam1961}. By slowly sweeping the detuning back and forth through resonance, one makes the magnetization flip repeatedly without loss, as long as the sweeping period is much shorter than $T_1$. At the flipping moment, transverse magnetization radiates a signal that can be captured by lock-in detection, in phase with the detuning oscillation.     

If the model of a single spin interacting with a driving field and a large reservoir generally does not work in NMR, we believe it correctly describes sparse impurity ions, continuously interacting with a large amount of matrix ions, off-resonant with the rf field. The corresponding kinetic equation theory, has been developed in the context of resonant optical excitation~\cite{kessel1988}. To extend the results to ARP, we observe that most of the time the magnetization is aligned along the static field $\textbf{H}_0$ and undergoes spin-lattice relaxation at rate $1/T_1$. During the flipping time, when spin-spin relaxation must be taken into account, long-lived locking results from $T_2$-lengthening rather than from thermodynamic equilibrium at spin temperature.


Following the lines of Ref.~\cite{kessel1988}, let us assign transverse relaxation to a fluctuating magnetic perturbation, whose amplitude $\Delta$ is uniformly distributed over a $\delta$-wide interval, such as $\delta\ll\gamma H_1$. The fluctuation correlation time is denoted $\tau_c$. According to Ref.~\cite{kessel1988}, the magnetization decays with the characteristic time:
\begin{equation}
\label{slowing_down}
\frac{1}{T(H_1)}=\frac{1}{T_1}+\frac{1}{T_2'}\frac{1}{1+(\gamma H_1\tau_c)^2},
\end{equation}
where $T_2'$ can be expressed as:
\begin{equation}
1/T_2'=\delta^2\tau_c
\end{equation}
and is related to zero-field $T_2$ by:
\begin{equation}
\label{T_2&T_2'}
T_2'=T_2\frac{3}{\zeta^2}\left[\frac{\zeta}{\arctan(\zeta)}-1\right],
\end{equation}    
where $\zeta=\sqrt{3}\;\delta\tau_c$. According to Eq. (\ref{T_2&T_2'}), $T_2'$ keeps close to $T_2$ over a wide range of variation of $\tau_c/T_2$, coinciding with $T_2$ in the white frequency noise limit where $\tau_c/T_2\ll 1$.
Hence, Eq.~\ref{slowing_down} expresses TR slowing down as $\gamma H_1\tau_c$ gets larger than unity. One may notice that the small amplitude perturbation condition $\delta\ll\gamma H_1$ is not enough to get rid of TR. 
           
If $\tau_c\gg\tau_f$, the transition frequency evolution is too slow to affect the final magnetization. Such a slow change only results in a shift of the moment when flipping takes place. If $\tau_c<\tau_f$, the spin is immune to frequency fluctuations provided $\tau_f\ll T(H_1)$. If $T(H_1)<T_1$, the condition $\tau_f\ll T(H_1)$ can be expressed as:
\begin{equation}
\label{eq:relaxation-free}
\delta^2/r\ll\gamma H_1\tau_c/\pi.
\end{equation}

This analysis fails when the magnetization is not initially aligned along $\textbf {H}_0$. Then the component orthogonal to $\textbf {H}_0$ relaxes with characteristic time T$_2$. The decay regime remains unchanged until the driving vector is dominated by the rf field. Then the relaxation might be slowed down  during the flipping time. However one must keep in mind the long duration of the adiabatic passage, supposed to be much longer than $\tau_f$. Hence the spin is subject to full relaxation, without slowing down, most of the time. 

In the latter context, let us examine the spin locking effect more precisely. We rely on Ref.~\cite{yodh1984}. Let us consider a spin at frequency $\omega_0+\Delta$ at time $t\approx0$. The magnetization component orthogonal to the driving vector $\mathbf{\Omega}=(\gamma H_1,0,\Delta)$, precesses around $\mathbf{\Omega}$ at angular velocity $\left[(\gamma H_1)^2+\Delta^2\right]^{1/2}$. With respect to spins at frequency $\omega_0$, the precession angle after a time interval $\tau$ differs by $\theta\approx\Delta^2\tau/(2\gamma H_1)$. Dephasing has been suspended over frequency interval $\Delta$ for a time $\tau$ if $\theta\ll\pi$. Identifying $\tau$ with $\tau_f$, one may conclude that, when $\tau_c<\tau_f$, the orthogonal magnetization component is not affected by frequency fluctuations if:
\begin{equation}
\label{eq:relaxation-free-transverse}
\delta^2/r\ll 2,
\end{equation}
a condition much more stringent than Eq.~\ref{eq:relaxation-free}, when $\gamma H_1\tau_c\gg 2\pi$.
       
\section{Experimental}\label{experiment} 
\begin{figure}
\includegraphics[width=7.1cm]{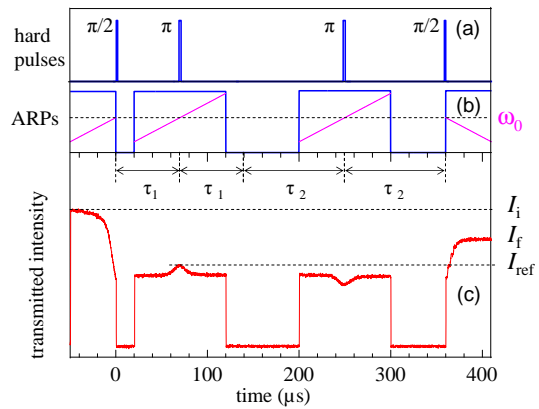}
\caption{(color online) Spin refocusing experiment for measuring $T_2$. (a) Equivalent hard pulse sequence, (b) ARP sequence, describing amplitude ($\gamma H_1=264$kHz) and frequency variations (chirp range: 6MHz) of the rf driving field, (c) transmitted optical intensity. After optical pumping the medium is transparent ($I_i$). The first AHP rotates the spins and restores the equilibrium transmission. The transmitted intensity raises to a maximum $I_\mathrm{ref}$ during the first AFP and behaves in the opposite way during the second AFP~\cite{lauro2011}. At time $2(\tau_1+\tau_2)$ after the end of the first AHP, the spins are phased back together. A second AHP converts the transverse magnetization back into a level population. The final transmitted intensity, $I_f$, reflects the spin coherence decay during the time interval $2(\tau_1+\tau_2)$. The ARP flipping time $\tau_f$ has been kept equal to 4.4$\mu s\ll 2(\tau_1+\tau_2)$}
\label{fig:refocusing_experiment}
\end{figure}
\subsection{Optically detected nuclear magnetic resonance}
The experiments are carried out in a 0.1 \% at. Tm$^{3+}$:YAG crystal. An external magnetic field lifts the nuclear spin degeneracy. The resulting four-level structure is comprised of two ground states and two excited states. We choose the same field orientation as in Ref.~\cite{louchet2007}.

The optical aspects of the setup have been described extensively in Refs.\cite{louchet2007} and \cite{lauro2009}. Briefly, the light beam, emerging from an extended cavity diode laser, is time-shaped by acousto-optic modulators, driven by a high sample-rate arbitrary waveform generator (Tektronix AWG 5004). The crystal is cooled down to 1.7 K in a liquid helium cryostat. The absorption depth of the $L=5$ mm-long sample is measured to be $\alpha L=0.6\pm0.02$. The static magnetic field is generated by superconducting coils and is set to about 0.5 T, which leads to a ground state splitting of 13.5 MHz. As in Ref.\cite{lauro2011}, the spin transition is resonantly driven by a rf magnetic field. Magnetic excitation is conveyed to the crystal by a 10 turn, 20 mm long, 10 mm in diameter, coil oriented along the light pulse wave vector. The crystal sits at the coil center. The rf signal, generated by the AWG, is fed to the coil through a pulsed amplifier (TOMCO BT00500-AlphaSA).

\begin{figure}
\includegraphics[width=7.1cm]{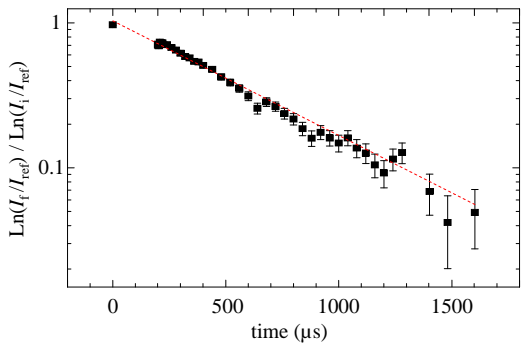}
\caption{(color online) Decay of the optical transmission at the end of the spin refocusing sequence as a function of $2(\tau_1+\tau_2)$ (see Fig.~\ref{fig:refocusing_experiment}). The exponential decay fit (red dashed line) leads to $T_2=550\pm$12 $\mu$s.} 
\label{fig:transverse_decay}
\end{figure}

The spin-lattice relaxation time $T_1$ reaches several seconds, as reported in a previous work~\cite{louchet2007} and by far exceeds all the other characteristic times. 

\subsection{Transverse relaxation measurement}
We measure $T_2$ by means of optically detected spin echo~\cite{shelby1978}. For that, we initialize the system by optically pumping the crystal into a single energy level of the nuclear spin. Pumping is achieved by a monochromatic laser, stabilized at 793 nm. Because of saturation effects, the pumping interval spreads over a few hundreds kHz. Since the optical transition is inhomogeneously broadened over $\approx$20 GHz, the preparation in a single state only impacts $\approx$10$^{-4}$ of the present Tm$^{3+}$ ions. With 0.1$\%$ at. Tm$^{3+}$ concentration and an irradiated volume radius of $\approx50\mu m$, the investigated sample contains less than 10$^{10}$ ions, which is worth noticing given the smallness of the optical oscillator strength (a few $10^{-8}$).  
 
A hard pulse spin echo sequence involves three evenly spaced pulses, with $\pi/2$, $\pi$, $\pi/2$ areas respectively~\cite{breiland1975}. The first $\pi/2$ pulse rotates the spins into the transverse plane, the $\pi$ pulse rephases the spins, and the last $\pi/2$ pulse converts the rephased transverse magnetization into an optically detectable level population. However the spin transition appears to be inhomogeneously broadened over $\Delta_\mathrm{in}$, such that $\Delta_\mathrm{in}/{2\pi}\gtrsim500$ kHz. Intense hard pulses, with duration smaller than  1 $\mu$s, would be needed to make all the prepared atoms contribute to the spin echo. Instead, we take advantage of the spin refocusing capabilities of a double ARP~\cite{lauro2011}, which strongly reduces the rf intensity requirements (see  Fig.~\ref{fig:refocusing_experiment}). After being rotated into the transverse plane by an adiabatic half passage (AHP), the spins depart from each other as they precess around $\textbf{H}_0$ at different speeds. Afterwards, they are refocused by a pair of adiabatic full passages (AFP). At the moment when they are aligned back together, a second AHP converts the magnetization back into a population difference that is monitored by optical transmission. The signal decay, as a function of the time interval between the two AHPs, is displayed in Fig.~\ref{fig:transverse_decay}. Experimental data are consistent with an exponential decay, in agreement with previous measurements on optical coherences in Tm$^{3+}$:YAG, performed under magnetic field~\cite{macfarlane1993}. We obtain $T_2=550\pm 12\mu s$, a little in excess of our previous measurement~\cite{lauro2011}, which may result from a slight tilt of the crystal with respect to $\textbf{H}_0$. The ARP flipping time, $\tau_f$, has been kept equal to 4.4 $\mu s\ll 2(\tau_1+\tau_2)$. Hence the spin freely evolves in the transverse plane most of the time.     
\begin{figure}
\includegraphics[width=7.1cm]{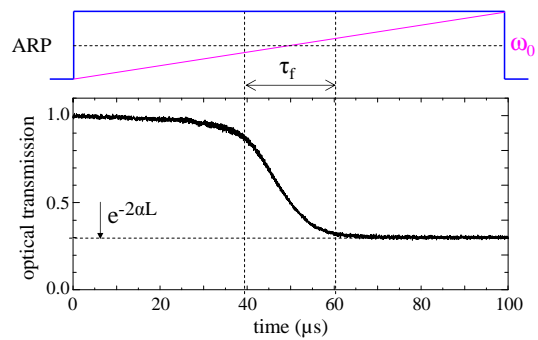}
\caption{(color online) Optical monitoring of a rf adiabatic rapid passage. Upper box: amplitude and frequency variations of the rf driving field. Lower box: transmission of the probe beam. At time $t=0$, the sample is transparent at the probe wavelength since the ground state of the probe transition has been emptied by optical pumping. Then optical transmission decreases to $e^{-2\alpha L}$ as the spins are flipped by the rf ARP. The flipping time is derived from the experimental parameter values: $\gamma H_1/(2\pi)=0.264$ MHz, $r/(2\pi)=40$ GHz/s. }
\label{fig:ARP}
\end{figure}	

\begin{figure}
\includegraphics[width=7.1cm]{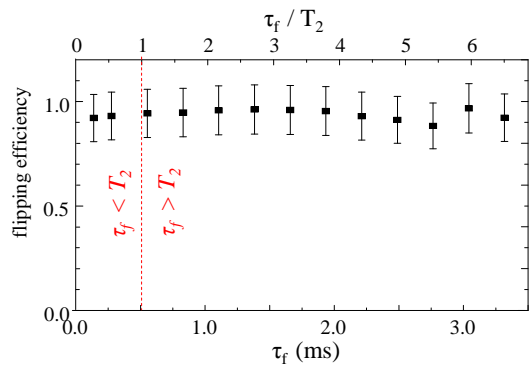}
\caption{(color online) Flipping efficiency as a function of flipping time. The sweeping range $\Delta_0$ and the Rabi frequency of the rf field are kept constant. The flipping efficiency is measured for different values of the chirp rate $r$. The flipping time $\tau_f$ varies from a small fraction of $T_2$ to more than 6 times $T_2$.}
\label{fig:flipping_efficiency}
\end{figure}
\subsection{Adiabatic passage at low flipping rate }	
Next, we turn to the measurement of flipping efficiency as a function of $\tau_f$. In the same way as before, we first optically pump the crystal into a single energy level of the nuclear spin. Then we apply a full ARP pulse that flips the atoms into their other spin level. The spin flip is monitored by optical transmission of a probe beam, as shown in Fig.~\ref{fig:ARP}. The finite duration of the ARP must be much larger than $\tau_f$, the flipping time, and than $\Delta_\mathrm{in}/r$, the sweeping time of the inhomogeneous width. Those two conditions are satisfied provided the rf field sweeping range $\Delta_0$ is much larger than $\gamma H_1$ and than $\Delta_\mathrm{in}$. With $\gamma H_1/(2\pi)=0.264$ MHz and $\Delta_0/(2\pi)=6$ MHz, both conditions are satisfied. By adjusting the chirp rate $r$, we are able to make $\tau_f$ vary from a fraction of $T_2$ to more than 3 ms. For large values of $\tau_f$, we only probe the level population at the beginning and at the end of the ARP. Indeed, too long exposition to the probe beam results in optical pumping that modifies the level population. The passage has been fully monitored for short $\tau_f$ values only, as displayed in Fig.~\ref{fig:ARP}   

All over the explored domain, the ARP condition, $(\gamma H_1)^2/r\gg 1$, is fulfilled. The results, displayed in Fig.~\ref{fig:flipping_efficiency}, show that the flipping efficiency remains close to unity over this interval. Hence TR appears to be inhibited by the ARP. In Tm$^{3+}$:YAG, the frequency fluctuations are mainly caused by the aluminium nuclear spins. Then, a fluctuation range value of a few kHz can be derived from Van Vleck formula~\cite{vanvleck1948}, by far smaller than $\gamma H_1$. With $\tau_c\approx 0.2$ ms~\cite{pascual}, it appears that $T_2'(\gamma H_1\tau_c)^2$ is about ten times larger than $T_1$. Hence, according to Eq.~\ref{slowing_down}, the magnetization decay is actually dominated by spin-lattice relaxation, all along the ARP process. 

\section{Conclusion}
Adiabatic passage on a nuclear spin transition can operate at slow rate, with a flipping time much larger than the transverse relaxation time T$_2$. We demonstrate this property experimentally in a low-concentration Tm$^{3+}$:YAG impurity-doped crystal, very different from the mono-atomic solids where similar features were observed in the early times of NMR. We relate this effect to spin-locking instead of spin temperature, as in those ancient investigations. According to a simple model, adiabatic passage is  robust to transition frequency fluctuations induced by the neighbor spins. 

However, this only works when the magnetization is initially aligned along the static field direction. Otherwise, the transverse component does not strongly interact with the driving field most of the time, and thus relaxes with characteristic time T$_2$. Therefore spin-locking during an adiabatic passage probably offers limited prospects as a dynamical decoupling technique.    
\begin{acknowledgments}
This research has been supported by European FP7-QuReP(STREP-247743) and FP7-CIPRIS(MC ITN-287252) projects, by the Agence Nationale de la Recherche ANR-09-BLAN-0333-03 project and by the Direction G\'en\'erale de l\'\ Armement.
\end{acknowledgments}


\begin{thebibliography}{41}
\expandafter\ifx\csname natexlab\endcsname\relax\def\natexlab#1{#1}\fi
\expandafter\ifx\csname bibnamefont\endcsname\relax
  \def\bibnamefont#1{#1}\fi
\expandafter\ifx\csname bibfnamefont\endcsname\relax
  \def\bibfnamefont#1{#1}\fi
\expandafter\ifx\csname citenamefont\endcsname\relax
  \def\citenamefont#1{#1}\fi
\expandafter\ifx\csname url\endcsname\relax
  \def\url#1{\texttt{#1}}\fi
\expandafter\ifx\csname urlprefix\endcsname\relax\def\urlprefix{URL }\fi
\providecommand{\bibinfo}[2]{#2}
\providecommand{\eprint}[2][]{\url{#2}}

\bibitem[{\citenamefont{Fleischhauer and Lukin}(2000)}]{fleischhauer2000}
\bibinfo{author}{\bibfnamefont{M.}~\bibnamefont{Fleischhauer}}
  \bibnamefont{and} \bibinfo{author}{\bibfnamefont{M.~D.} \bibnamefont{Lukin}},
  \bibinfo{journal}{Phys. Rev. Lett.} \textbf{\bibinfo{volume}{84}},
  \bibinfo{pages}{5094} (\bibinfo{year}{2000}).

\bibitem[{\citenamefont{Moiseev and Kr\"oll}(2001)}]{moiseev2001}
\bibinfo{author}{\bibfnamefont{S.~A.} \bibnamefont{Moiseev}} \bibnamefont{and}
  \bibinfo{author}{\bibfnamefont{S.}~\bibnamefont{Kr\"oll}},
  \bibinfo{journal}{Phys. Rev. Lett.} \textbf{\bibinfo{volume}{87}},
  \bibinfo{pages}{173601} (\bibinfo{year}{2001}).

\bibitem[{\citenamefont{Nilsson and Kr{\"o}ll}(2005)}]{nilsson2005}
\bibinfo{author}{\bibfnamefont{M.}~\bibnamefont{Nilsson}} \bibnamefont{and}
  \bibinfo{author}{\bibfnamefont{S.}~\bibnamefont{Kr{\"o}ll}},
  \bibinfo{journal}{Opt. Commun.} \textbf{\bibinfo{volume}{247}},
  \bibinfo{pages}{393} (\bibinfo{year}{2005}).

\bibitem[{\citenamefont{Lauro et~al.}(2009{\natexlab{a}})\citenamefont{Lauro,
  Chaneli\`ere, and Le~Gou\"et}}]{lauro2009a}
\bibinfo{author}{\bibfnamefont{R.}~\bibnamefont{Lauro}},
  \bibinfo{author}{\bibfnamefont{T.}~\bibnamefont{Chaneli\`ere}},
  \bibnamefont{and} \bibinfo{author}{\bibfnamefont{J.-L.}
  \bibnamefont{Le~Gou\"et}}, \bibinfo{journal}{Phys. Rev. A}
  \textbf{\bibinfo{volume}{79}}, \bibinfo{pages}{053801}
  (\bibinfo{year}{2009}{\natexlab{a}}).

\bibitem[{\citenamefont{Afzelius et~al.}(2009)\citenamefont{Afzelius, Simon,
  de~Riedmatten, and Gisin}}]{afzelius2009}
\bibinfo{author}{\bibfnamefont{M.}~\bibnamefont{Afzelius}},
  \bibinfo{author}{\bibfnamefont{C.}~\bibnamefont{Simon}},
  \bibinfo{author}{\bibfnamefont{H.}~\bibnamefont{de~Riedmatten}},
  \bibnamefont{and} \bibinfo{author}{\bibfnamefont{N.}~\bibnamefont{Gisin}},
  \bibinfo{journal}{Phys. Rev. A} \textbf{\bibinfo{volume}{79}},
  \bibinfo{pages}{052329} (\bibinfo{year}{2009}).

\bibitem[{\citenamefont{Afzelius et~al.}(2010)\citenamefont{Afzelius, Usmani,
  Amari, Lauritzen, Walther, Simon, Sangouard, Min\'a\ifmmode~\check{r}\else
  \v{r}\fi{}, de~Riedmatten, Gisin et~al.}}]{afzelius2010}
\bibinfo{author}{\bibfnamefont{M.}~\bibnamefont{Afzelius}},
  \bibinfo{author}{\bibfnamefont{I.}~\bibnamefont{Usmani}},
  \bibinfo{author}{\bibfnamefont{A.}~\bibnamefont{Amari}},
  \bibinfo{author}{\bibfnamefont{B.}~\bibnamefont{Lauritzen}},
  \bibinfo{author}{\bibfnamefont{A.}~\bibnamefont{Walther}},
  \bibinfo{author}{\bibfnamefont{C.}~\bibnamefont{Simon}},
  \bibinfo{author}{\bibfnamefont{N.}~\bibnamefont{Sangouard}},
  \bibinfo{author}{\bibfnamefont{J.}~\bibnamefont{Min\'a\ifmmode~\check{r}\else
  \v{r}\fi{}}},
  \bibinfo{author}{\bibfnamefont{H.}~\bibnamefont{de~Riedmatten}},
  \bibinfo{author}{\bibfnamefont{N.}~\bibnamefont{Gisin}},
  \bibnamefont{et~al.}, \bibinfo{journal}{Phys. Rev. Lett.}
  \textbf{\bibinfo{volume}{104}}, \bibinfo{pages}{040503}
  (\bibinfo{year}{2010}).

\bibitem[{\citenamefont{Turukhin et~al.}(2001)\citenamefont{Turukhin,
  Sudarshanam, Shahriar, Musser, Ham, and Hemmer}}]{turukhin2001}
\bibinfo{author}{\bibfnamefont{A.~V.} \bibnamefont{Turukhin}},
  \bibinfo{author}{\bibfnamefont{V.~S.} \bibnamefont{Sudarshanam}},
  \bibinfo{author}{\bibfnamefont{M.~S.} \bibnamefont{Shahriar}},
  \bibinfo{author}{\bibfnamefont{J.~A.} \bibnamefont{Musser}},
  \bibinfo{author}{\bibfnamefont{B.~S.} \bibnamefont{Ham}}, \bibnamefont{and}
  \bibinfo{author}{\bibfnamefont{P.~R.} \bibnamefont{Hemmer}},
  \bibinfo{journal}{Phys. Rev. Lett.} \textbf{\bibinfo{volume}{88}},
  \bibinfo{pages}{023602} (\bibinfo{year}{2001}).

\bibitem[{\citenamefont{Longdell et~al.}(2005)\citenamefont{Longdell, Fraval,
  Sellars, and Manson}}]{longdell2005}
\bibinfo{author}{\bibfnamefont{J.~J.} \bibnamefont{Longdell}},
  \bibinfo{author}{\bibfnamefont{E.}~\bibnamefont{Fraval}},
  \bibinfo{author}{\bibfnamefont{M.~J.} \bibnamefont{Sellars}},
  \bibnamefont{and} \bibinfo{author}{\bibfnamefont{N.~B.}
  \bibnamefont{Manson}}, \bibinfo{journal}{Phys. Rev. Lett.}
  \textbf{\bibinfo{volume}{95}}, \bibinfo{pages}{063601}
  (\bibinfo{year}{2005}).

\bibitem[{\citenamefont{Fraval et~al.}(2004)\citenamefont{Fraval, Sellars, and
  Longdell}}]{fraval2004}
\bibinfo{author}{\bibfnamefont{E.}~\bibnamefont{Fraval}},
  \bibinfo{author}{\bibfnamefont{M.~J.} \bibnamefont{Sellars}},
  \bibnamefont{and} \bibinfo{author}{\bibfnamefont{J.~J.}
  \bibnamefont{Longdell}}, \bibinfo{journal}{Phys. Rev. Lett.}
  \textbf{\bibinfo{volume}{92}}, \bibinfo{pages}{077601}
  (\bibinfo{year}{2004}).

\bibitem[{\citenamefont{Lovri\ifmmode~\acute{c}\else \'{c}\fi{}
  et~al.}(2011)\citenamefont{Lovri\ifmmode~\acute{c}\else \'{c}\fi{},
  Glasenapp, Suter, Tumino, Ferrier, Goldner, Sabooni, Rippe, and
  Kr\"oll}}]{lovric2011}
\bibinfo{author}{\bibfnamefont{M.}~\bibnamefont{Lovri\ifmmode~\acute{c}\else
  \'{c}\fi{}}}, \bibinfo{author}{\bibfnamefont{P.}~\bibnamefont{Glasenapp}},
  \bibinfo{author}{\bibfnamefont{D.}~\bibnamefont{Suter}},
  \bibinfo{author}{\bibfnamefont{B.}~\bibnamefont{Tumino}},
  \bibinfo{author}{\bibfnamefont{A.}~\bibnamefont{Ferrier}},
  \bibinfo{author}{\bibfnamefont{P.}~\bibnamefont{Goldner}},
  \bibinfo{author}{\bibfnamefont{M.}~\bibnamefont{Sabooni}},
  \bibinfo{author}{\bibfnamefont{L.}~\bibnamefont{Rippe}}, \bibnamefont{and}
  \bibinfo{author}{\bibfnamefont{S.}~\bibnamefont{Kr\"oll}},
  \bibinfo{journal}{Phys. Rev. B} \textbf{\bibinfo{volume}{84}},
  \bibinfo{pages}{104417} (\bibinfo{year}{2011}).

\bibitem[{\citenamefont{Carr and Purcell}(1954)}]{carr1954}
\bibinfo{author}{\bibfnamefont{H.~Y.} \bibnamefont{Carr}} \bibnamefont{and}
  \bibinfo{author}{\bibfnamefont{E.~M.} \bibnamefont{Purcell}},
  \bibinfo{journal}{Phys. Rev.} \textbf{\bibinfo{volume}{94}},
  \bibinfo{pages}{630} (\bibinfo{year}{1954}).

\bibitem[{\citenamefont{Meiboom and Gill}(1958)}]{meiboom1958}
\bibinfo{author}{\bibfnamefont{S.}~\bibnamefont{Meiboom}} \bibnamefont{and}
  \bibinfo{author}{\bibfnamefont{D.}~\bibnamefont{Gill}},
  \bibinfo{journal}{Rev. Scient. Instr.} \textbf{\bibinfo{volume}{29}},
  \bibinfo{pages}{688} (\bibinfo{year}{1958}).

\bibitem[{\citenamefont{Fraval et~al.}(2005)\citenamefont{Fraval, Sellars, and
  Longdell}}]{fraval2005}
\bibinfo{author}{\bibfnamefont{E.}~\bibnamefont{Fraval}},
  \bibinfo{author}{\bibfnamefont{M.~J.} \bibnamefont{Sellars}},
  \bibnamefont{and} \bibinfo{author}{\bibfnamefont{J.~J.}
  \bibnamefont{Longdell}}, \bibinfo{journal}{Phys. Rev. Lett.}
  \textbf{\bibinfo{volume}{95}}, \bibinfo{pages}{030506}
  (\bibinfo{year}{2005}).

\bibitem[{\citenamefont{Solomon}(1959)}]{solomon1959}
\bibinfo{author}{\bibfnamefont{I.}~\bibnamefont{Solomon}},
  \bibinfo{journal}{Comptes Rendus Acad. Sci. paris}
  \textbf{\bibinfo{volume}{248}}, \bibinfo{pages}{92} (\bibinfo{year}{1959}).

\bibitem[{\citenamefont{Hartmann and Hahn}(1962)}]{hartmann1962}
\bibinfo{author}{\bibfnamefont{S.~R.} \bibnamefont{Hartmann}} \bibnamefont{and}
  \bibinfo{author}{\bibfnamefont{E.~L.} \bibnamefont{Hahn}},
  \bibinfo{journal}{Phys. Rev.} \textbf{\bibinfo{volume}{128}},
  \bibinfo{pages}{2042} (\bibinfo{year}{1962}).

\bibitem[{\citenamefont{DeVoe and Brewer}(1983)}]{devoe1983}
\bibinfo{author}{\bibfnamefont{R.~G.} \bibnamefont{DeVoe}} \bibnamefont{and}
  \bibinfo{author}{\bibfnamefont{R.~G.} \bibnamefont{Brewer}},
  \bibinfo{journal}{Phys. Rev. Lett.} \textbf{\bibinfo{volume}{50}},
  \bibinfo{pages}{1269} (\bibinfo{year}{1983}).

\bibitem[{\citenamefont{Pryde et~al.}(2004)\citenamefont{Pryde, Sellars, and
  Manson}}]{pryde2004}
\bibinfo{author}{\bibfnamefont{G.~J.} \bibnamefont{Pryde}},
  \bibinfo{author}{\bibfnamefont{M.~J.} \bibnamefont{Sellars}},
  \bibnamefont{and} \bibinfo{author}{\bibfnamefont{N.~B.}
  \bibnamefont{Manson}}, \bibinfo{journal}{Phys. Rev. B}
  \textbf{\bibinfo{volume}{69}}, \bibinfo{pages}{075107}
  (\bibinfo{year}{2004}).

\bibitem[{\citenamefont{Yodh et~al.}(1984)\citenamefont{Yodh, Golub, Carlson,
  and Mossberg}}]{yodh1984}
\bibinfo{author}{\bibfnamefont{A.~G.} \bibnamefont{Yodh}},
  \bibinfo{author}{\bibfnamefont{J.}~\bibnamefont{Golub}},
  \bibinfo{author}{\bibfnamefont{N.~W.} \bibnamefont{Carlson}},
  \bibnamefont{and} \bibinfo{author}{\bibfnamefont{T.~W.}
  \bibnamefont{Mossberg}}, \bibinfo{journal}{Phys. Rev. Lett.}
  \textbf{\bibinfo{volume}{53}}, \bibinfo{pages}{659} (\bibinfo{year}{1984}).

\bibitem[{\citenamefont{Carlson et~al.}(1984)\citenamefont{Carlson, Babbitt,
  BAI, and Mossberg}}]{carlson1984}
\bibinfo{author}{\bibfnamefont{N.}~\bibnamefont{Carlson}},
  \bibinfo{author}{\bibfnamefont{W.}~\bibnamefont{Babbitt}},
  \bibinfo{author}{\bibfnamefont{Y.}~\bibnamefont{BAI}}, \bibnamefont{and}
  \bibinfo{author}{\bibfnamefont{T.}~\bibnamefont{Mossberg}},
  \bibinfo{journal}{Opt. Lett.} \textbf{\bibinfo{volume}{9}},
  \bibinfo{pages}{232} (\bibinfo{year}{1984}).

\bibitem[{\citenamefont{Tch\'enio et~al.}(1988)\citenamefont{Tch\'enio,
  D\'ebarre, Keller, and Le~Gou\"et}}]{tchenio1988}
\bibinfo{author}{\bibfnamefont{P.}~\bibnamefont{Tch\'enio}},
  \bibinfo{author}{\bibfnamefont{A.}~\bibnamefont{D\'ebarre}},
  \bibinfo{author}{\bibfnamefont{J.-C.} \bibnamefont{Keller}},
  \bibnamefont{and} \bibinfo{author}{\bibfnamefont{J.-L.}
  \bibnamefont{Le~Gou\"et}}, \bibinfo{journal}{Phys. Rev. A}
  \textbf{\bibinfo{volume}{38}}, \bibinfo{pages}{5235} (\bibinfo{year}{1988}).

\bibitem[{\citenamefont{de~Seze et~al.}(2006)\citenamefont{de~Seze, Louchet,
  Crozatier, Lorger\'e, Bretenaker, Le~Gou\"et, Guillot-No\"el, and
  Goldner}}]{deSeze2006}
\bibinfo{author}{\bibfnamefont{F.}~\bibnamefont{de~Seze}},
  \bibinfo{author}{\bibfnamefont{A.}~\bibnamefont{Louchet}},
  \bibinfo{author}{\bibfnamefont{V.}~\bibnamefont{Crozatier}},
  \bibinfo{author}{\bibfnamefont{I.}~\bibnamefont{Lorger\'e}},
  \bibinfo{author}{\bibfnamefont{F.}~\bibnamefont{Bretenaker}},
  \bibinfo{author}{\bibfnamefont{J.-L.} \bibnamefont{Le~Gou\"et}},
  \bibinfo{author}{\bibfnamefont{O.}~\bibnamefont{Guillot-No\"el}},
  \bibnamefont{and} \bibinfo{author}{\bibfnamefont{P.}~\bibnamefont{Goldner}},
  \bibinfo{journal}{Phys. Rev. B} \textbf{\bibinfo{volume}{73}},
  \bibinfo{pages}{085112} (\bibinfo{year}{2006}).

\bibitem[{\citenamefont{Louchet et~al.}(2007)\citenamefont{Louchet, Habib,
  Crozatier, Lorger\'e, Goldfarb, Bretenaker, Le~Gou\"et, Guillot-No\"el, and
  Goldner}}]{louchet2007}
\bibinfo{author}{\bibfnamefont{A.}~\bibnamefont{Louchet}},
  \bibinfo{author}{\bibfnamefont{J.~S.} \bibnamefont{Habib}},
  \bibinfo{author}{\bibfnamefont{V.}~\bibnamefont{Crozatier}},
  \bibinfo{author}{\bibfnamefont{I.}~\bibnamefont{Lorger\'e}},
  \bibinfo{author}{\bibfnamefont{F.}~\bibnamefont{Goldfarb}},
  \bibinfo{author}{\bibfnamefont{F.}~\bibnamefont{Bretenaker}},
  \bibinfo{author}{\bibfnamefont{J.-L.} \bibnamefont{Le~Gou\"et}},
  \bibinfo{author}{\bibfnamefont{O.}~\bibnamefont{Guillot-No\"el}},
  \bibnamefont{and} \bibinfo{author}{\bibfnamefont{P.}~\bibnamefont{Goldner}},
  \bibinfo{journal}{Phys. Rev. B} \textbf{\bibinfo{volume}{75}},
  \bibinfo{pages}{035131} (\bibinfo{year}{2007}).

\bibitem[{\citenamefont{Louchet et~al.}(2008)\citenamefont{Louchet, Le~Du,
  Bretenaker, Chaneli\`ere, Goldfarb, Lorger\'e, Le~Gou\"et, Guillot-No\"el,
  and Goldner}}]{louchet2008}
\bibinfo{author}{\bibfnamefont{A.}~\bibnamefont{Louchet}},
  \bibinfo{author}{\bibfnamefont{Y.}~\bibnamefont{Le~Du}},
  \bibinfo{author}{\bibfnamefont{F.}~\bibnamefont{Bretenaker}},
  \bibinfo{author}{\bibfnamefont{T.}~\bibnamefont{Chaneli\`ere}},
  \bibinfo{author}{\bibfnamefont{F.}~\bibnamefont{Goldfarb}},
  \bibinfo{author}{\bibfnamefont{I.}~\bibnamefont{Lorger\'e}},
  \bibinfo{author}{\bibfnamefont{J.-L.} \bibnamefont{Le~Gou\"et}},
  \bibinfo{author}{\bibfnamefont{O.}~\bibnamefont{Guillot-No\"el}},
  \bibnamefont{and} \bibinfo{author}{\bibfnamefont{P.}~\bibnamefont{Goldner}},
  \bibinfo{journal}{Phys. Rev. B} \textbf{\bibinfo{volume}{77}},
  \bibinfo{pages}{195110} (\bibinfo{year}{2008}).

\bibitem[{\citenamefont{Lauro et~al.}(2009{\natexlab{b}})\citenamefont{Lauro,
  Chaneli\`{e}re, and Le~Gou\"{e}t}}]{lauro2009}
\bibinfo{author}{\bibfnamefont{R.}~\bibnamefont{Lauro}},
  \bibinfo{author}{\bibfnamefont{T.}~\bibnamefont{Chaneli\`{e}re}},
  \bibnamefont{and} \bibinfo{author}{\bibfnamefont{J.~L.}
  \bibnamefont{Le~Gou\"{e}t}}, \bibinfo{journal}{Phys. Rev. A}
  \textbf{\bibinfo{volume}{79}}, \bibinfo{pages}{063844}
  (\bibinfo{year}{2009}{\natexlab{b}}).

\bibitem[{\citenamefont{Chanelière et~al.}(2010)\citenamefont{Chanelière,
  Ruggiero, Bonarota, Afzelius, and Le~Gouët}}]{chaneliere2010}
\bibinfo{author}{\bibfnamefont{T.}~\bibnamefont{Chanelière}},
  \bibinfo{author}{\bibfnamefont{J.}~\bibnamefont{Ruggiero}},
  \bibinfo{author}{\bibfnamefont{M.}~\bibnamefont{Bonarota}},
  \bibinfo{author}{\bibfnamefont{M.}~\bibnamefont{Afzelius}}, \bibnamefont{and}
  \bibinfo{author}{\bibfnamefont{J.-L.} \bibnamefont{Le~Gouët}},
  \bibinfo{journal}{N. J. Phys.} \textbf{\bibinfo{volume}{12}},
  \bibinfo{pages}{023025} (\bibinfo{year}{2010}).

\bibitem[{\citenamefont{Bonarota et~al.}(2010)\citenamefont{Bonarota, Ruggiero,
  Le~Gouët, and Chanelière}}]{bonarota2010}
\bibinfo{author}{\bibfnamefont{M.}~\bibnamefont{Bonarota}},
  \bibinfo{author}{\bibfnamefont{J.}~\bibnamefont{Ruggiero}},
  \bibinfo{author}{\bibfnamefont{J.-L.} \bibnamefont{Le~Gouët}},
  \bibnamefont{and}
  \bibinfo{author}{\bibfnamefont{T.}~\bibnamefont{Chanelière}},
  \bibinfo{journal}{Phys. Rev. A} \textbf{\bibinfo{volume}{81}},
  \bibinfo{pages}{033803} (\bibinfo{year}{2010}).

\bibitem[{\citenamefont{Lauro et~al.}(2011)\citenamefont{Lauro, Chaneli\`ere,
  and Le~Gou\"et}}]{lauro2011}
\bibinfo{author}{\bibfnamefont{R.}~\bibnamefont{Lauro}},
  \bibinfo{author}{\bibfnamefont{T.}~\bibnamefont{Chaneli\`ere}},
  \bibnamefont{and} \bibinfo{author}{\bibfnamefont{J.-L.}
  \bibnamefont{Le~Gou\"et}}, \bibinfo{journal}{Phys. Rev. B}
  \textbf{\bibinfo{volume}{83}}, \bibinfo{pages}{035124}
  (\bibinfo{year}{2011}).

\bibitem[{\citenamefont{Bonarota et~al.}(2011)\citenamefont{Bonarota, Le~Gouët,
  and Chanelière}}]{bonarota2011}
\bibinfo{author}{\bibfnamefont{M.}~\bibnamefont{Bonarota}},
  \bibinfo{author}{\bibfnamefont{J.-L.} \bibnamefont{Le~Gouët}},
  \bibnamefont{and}
  \bibinfo{author}{\bibfnamefont{T.}~\bibnamefont{Chanelière}},
  \bibinfo{journal}{New J. Phys.} \textbf{\bibinfo{volume}{13}},
  \bibinfo{pages}{013013} (\bibinfo{year}{2011}).

\bibitem[{\citenamefont{Redfield}(1955)}]{redfield1955}
\bibinfo{author}{\bibfnamefont{A.~G.} \bibnamefont{Redfield}},
  \bibinfo{journal}{Phys. Rev.} \textbf{\bibinfo{volume}{98}},
  \bibinfo{pages}{1787} (\bibinfo{year}{1955}).

\bibitem[{\citenamefont{Abragam}(1961)}]{abragam1961}
\bibinfo{author}{\bibfnamefont{A.}~\bibnamefont{Abragam}},
  \emph{\bibinfo{title}{Principles of Nuclear Magnetism}}
  (\bibinfo{publisher}{Oxford Univ. Press, Oxford, England},
  \bibinfo{year}{1961}).

\bibitem[{\citenamefont{Slichter and Holton}(1961)}]{slichter1961}
\bibinfo{author}{\bibfnamefont{C.~P.} \bibnamefont{Slichter}} \bibnamefont{and}
  \bibinfo{author}{\bibfnamefont{W.~C.} \bibnamefont{Holton}},
  \bibinfo{journal}{Phys. Rev.} \textbf{\bibinfo{volume}{122}},
  \bibinfo{pages}{1701} (\bibinfo{year}{1961}).

\bibitem[{\citenamefont{Solomon and Ezratty}(1962)}]{solomon1962}
\bibinfo{author}{\bibfnamefont{I.}~\bibnamefont{Solomon}} \bibnamefont{and}
  \bibinfo{author}{\bibfnamefont{J.}~\bibnamefont{Ezratty}},
  \bibinfo{journal}{Phys. Rev.} \textbf{\bibinfo{volume}{127}},
  \bibinfo{pages}{78} (\bibinfo{year}{1962}).

\bibitem[{\citenamefont{Bloch}(1946)}]{bloch1946}
\bibinfo{author}{\bibfnamefont{F.}~\bibnamefont{Bloch}},
  \bibinfo{journal}{Phys. Rev.} \textbf{\bibinfo{volume}{70}},
  \bibinfo{pages}{460} (\bibinfo{year}{1946}).

\bibitem[{\citenamefont{Schenzle et~al.}(1984)\citenamefont{Schenzle,
  Mitsunaga, DeVoe, and Brewer}}]{schenzle1984}
\bibinfo{author}{\bibfnamefont{A.}~\bibnamefont{Schenzle}},
  \bibinfo{author}{\bibfnamefont{M.}~\bibnamefont{Mitsunaga}},
  \bibinfo{author}{\bibfnamefont{R.~G.} \bibnamefont{DeVoe}}, \bibnamefont{and}
  \bibinfo{author}{\bibfnamefont{R.~G.} \bibnamefont{Brewer}},
  \bibinfo{journal}{Phys. Rev. A} \textbf{\bibinfo{volume}{30}},
  \bibinfo{pages}{325} (\bibinfo{year}{1984}).

\bibitem[{\citenamefont{Kessel et~al.}(1988)\citenamefont{Kessel, R.P., and
  Eksin}}]{kessel1988}
\bibinfo{author}{\bibfnamefont{A.}~\bibnamefont{Kessel}},
  \bibinfo{author}{\bibfnamefont{S.}~\bibnamefont{R.P.}}, \bibnamefont{and}
  \bibinfo{author}{\bibfnamefont{L.}~\bibnamefont{Eksin}},
  \bibinfo{journal}{Sov. Phys. JETP} \textbf{\bibinfo{volume}{67}},
  \bibinfo{pages}{2071} (\bibinfo{year}{1988}), \bibinfo{note}{($Zh. Eksp.
  Teor. Fiz.\textbf{94}, 202-215 (1988)$)}.

\bibitem[{\citenamefont{Lacour et~al.}(2007)\citenamefont{Lacour, Gu\'erin,
  Yatsenko, Vitanov, and Jauslin}}]{lacour2007}
\bibinfo{author}{\bibfnamefont{X.}~\bibnamefont{Lacour}},
  \bibinfo{author}{\bibfnamefont{S.}~\bibnamefont{Gu\'erin}},
  \bibinfo{author}{\bibfnamefont{L.~P.} \bibnamefont{Yatsenko}},
  \bibinfo{author}{\bibfnamefont{N.~V.} \bibnamefont{Vitanov}},
  \bibnamefont{and} \bibinfo{author}{\bibfnamefont{H.~R.}
  \bibnamefont{Jauslin}}, \bibinfo{journal}{Phys. Rev. A}
  \textbf{\bibinfo{volume}{75}}, \bibinfo{pages}{033417}
  (\bibinfo{year}{2007}).

\bibitem[{\citenamefont{Shelby et~al.}(1978)\citenamefont{Shelby, Yannoni, and
  Macfarlane}}]{shelby1978}
\bibinfo{author}{\bibfnamefont{R.~M.} \bibnamefont{Shelby}},
  \bibinfo{author}{\bibfnamefont{C.~S.} \bibnamefont{Yannoni}},
  \bibnamefont{and} \bibinfo{author}{\bibfnamefont{R.~M.}
  \bibnamefont{Macfarlane}}, \bibinfo{journal}{Phys. Rev. Lett.}
  \textbf{\bibinfo{volume}{41}}, \bibinfo{pages}{1739} (\bibinfo{year}{1978}).

\bibitem[{\citenamefont{Breiland et~al.}(1975)\citenamefont{Breiland, Brenner,
  and Harris}}]{breiland1975}
\bibinfo{author}{\bibfnamefont{W.~G.} \bibnamefont{Breiland}},
  \bibinfo{author}{\bibfnamefont{H.~C.} \bibnamefont{Brenner}},
  \bibnamefont{and} \bibinfo{author}{\bibfnamefont{C.~B.}
  \bibnamefont{Harris}}, \bibinfo{journal}{J. Chem. Phys.}
  \textbf{\bibinfo{volume}{62}}, \bibinfo{pages}{3458} (\bibinfo{year}{1975}).

\bibitem[{\citenamefont{Macfarlane}(1993)}]{macfarlane1993}
\bibinfo{author}{\bibfnamefont{R.~M.} \bibnamefont{Macfarlane}},
  \bibinfo{journal}{Opt. Lett.} \textbf{\bibinfo{volume}{18}},
  \bibinfo{pages}{1958} (\bibinfo{year}{1993}).

\bibitem[{\citenamefont{Van~Vleck}(1948)}]{vanvleck1948}
\bibinfo{author}{\bibfnamefont{J.~H.} \bibnamefont{Van~Vleck}},
  \bibinfo{journal}{Phys. Rev.} \textbf{\bibinfo{volume}{74}},
  \bibinfo{pages}{1168} (\bibinfo{year}{1948}).

\bibitem[{\citenamefont{Pascual~Winter et~al.}()\citenamefont{Pascual~Winter,
  Tongning, Louchet-Chauvet, Chaneli\`ere, and Le~Gou\"et}}]{pascual}
\bibinfo{author}{\bibfnamefont{M.~F.} \bibnamefont{Pascual~Winter}},
  \bibinfo{author}{\bibfnamefont{R.~C.} \bibnamefont{Tongning}},
  \bibinfo{author}{\bibfnamefont{A.}~\bibnamefont{Louchet-Chauvet}},
  \bibinfo{author}{\bibfnamefont{T.}~\bibnamefont{Chaneli\`ere}},
  \bibnamefont{and} \bibinfo{author}{\bibfnamefont{J.-L.}
  \bibnamefont{Le~Gou\"et}}, \bibinfo{note}{to be published}.

\end{thebibliography}
\end{document}